*Title:*  What is the Potential Impact of the IsoDAR Cyclotron on Radioisotope Production: A Review

*Authors:*    Loyd H. Waites[1]

Jose R. Alonso[1]

Roger Barlow[2]

Janet M. Conrad[1]

[1]    Physics Department, Massachusetts Institute of Technology, 77 Massachusetts Avenue, Cambridge, MA 02139, USA

[2]    School of Computing and Engineering, The University of Huddersfield, Queensgate, Huddersfield, HD1 3DH, UK

*Disclaimers*:  None, no conflicts of interest

*Corresponding Author*:  Loyd Hoyt Waites

Address:       26-540, MIT, 77 Massachusetts Avenue, Cambridge, MA 02139, USA

Telephone    +01 (617) 324 6281,

Fax:             +01 (617) 253 0111

e-mail          lwaites@mit.edu

**ABSTRACT**

The IsoDAR collaboration is developing a high-current cyclotron for a neutrino search experiment. Designed to deliver 10 mA of 60 MeV protons, the current and power of this cyclotron far exceed those of existing accelerators, opening new possibilities for the production of radiopharmaceutical isotopes, producing very high-activity samples in very short times. The cyclotron can also be easily configured to deliver ions other than protons including 1 mA of alpha particles at 240 MeV: this flexibility gives a broad reach into new areas of isotope production. We explain how IsoDAR overcomes the beam limits of commercial cyclotrons, and how it could represent the next step in isotope production rates.

# I  INTRODUCTION

Radiopharmaceutical isotopes are widely used in medical practice, for both imaging and therapy. Applications range widely but each begins with the creation of the artificial unstable isotope. This can be done in a reactor (though many are now being phased out) or using an accelerator, most always a cyclotron.

Such cyclotrons typically [1] accelerate a maximum of 2 mA of negative ion of hydrogen (H⁻) to 30 MeV, while the largest commercial machines from IBA [2] and Best [3] produce about 1 mA at 70 MeV. These energies are appropriate, covering the optimum production energies for a large number of the most desirable radioisotopes. The 60-MeV IsoDAR cyclotron increases the available proton current to 10 mA. This higher current provides a two-fold advantage: it enables a higher production rate of established medical isotopes, and it opens the possible deployment of isotopes with small production cross sections or long half-lives.

The IsoDAR experiment is designed to place a powerful neutrino source in close proximity to a large (on the order of 1000 cubic meters) liquid scintillator detector (specifically: KamLAND, in Japan's Kamioka Observatory) as a definitive test for the existence of sterile neutrinos [4]. The 600 kW of protons from the cyclotron strike a beryllium target to produce neutrons that flood a $^7$Li-containing sleeve generating $^8$Li, whose decay in turn produces the desired electron antineutrinos with a flux equivalent to that of a 2 petabecquerel (50 kiloCurie) beta-decay source.

A key to the increase in maximum current is the use of $H_2^+$, rather than H⁻ or protons, as the accelerated particle. A major reason for the current limit in high-current cyclotrons is "space charge" or the mutual electrostatic repulsion of particles in the beam that causes the beam to grow in size, making it more difficult to contain and prevent beam losses. The $H_2^+$ ion has one charge but contains two protons, reducing the total charge of the beam, hence the space-charge repulsion. In addition, the greater mass of the $H_2^+$ ion increases the inertia and slows the beam growth from the repulsive charge. As we will see in Sections II G, stripping extraction, a key feature of H⁻ cyclotrons, is also possible with $H_2^+$ ions. This choice of beam particle means that the cyclotron also has the flexibility to accelerate other ions with the same charge-to-mass ratio, such as deuterons, alpha particles or $C^{6+}$, opening possibilities, discussed in section III B, for generating isotopes not accessible with proton beams.

*Running Title:*  IsoDAR Cyclotron

There are many possible uses of such a high current source, such as $^{225}$Ac from natural thorium targets, or long-lived generator parents (e.g. $^{68}$Ge parent ($t_{1/2}$= 270 d) of the Ge/Ga generator), possibly even $^{148}$Gd as a nuclear battery replacement for $^{238}$Pu. These are discussed in section III.

The beam power exceeds the capabilities of present-day isotope production targets but possible obstacles (such as heat dissipation, the large facility footprint, or radiation shielding) we do not see as serious impediments to the development of higher capacity targets to make use of a high intensity beam. Alternatively, techniques for splitting the beam amongst many targets are discussed in section III A.

## II  THE IsoDAR CYCLOTRON

The IsoDAR design [5] is a compact cyclotron: these cyclotrons are the workhorse of the isotope industry. The rigorous demands of this field have led to mature designs, well-understood costs, and excellent operational reliability. We will lead the reader through how cyclotrons work, what their limits are and how the IsoDAR cyclotron overcomes these limits [6].

### A.    The Ion Source

State-of-the-art isotope cyclotrons inject beam from an external ion source (producing typically 5 to 10 mA of H$^-$ ions) placed above or below the cyclotron, with a short beam line running along the central axis of the magnet (perpendicular to the plane of the magnetic field). The source is held at a high voltage, typically 30 kV, providing the initial energy for the continuous beam. Ions are deflected into the midplane by a spiral inflector and directed to the first accelerating electrode of the cyclotron RF system (see Fig. 1).  IsoDAR follows the same scheme, except that $H_2^+$ ions are used rather than H$^-$.

In an ion source, typically, high-power microwaves (~2.5 GHz) drive a plasma discharge that removes an electron from the $H_2$ molecule, leaving $H_2^+$. Subsequent collisions may dissociate the $H_2^+$ ion into protons. Both species are extracted from the plasma by a high-voltage electrode and formed into a beam. The desired species is selected by an analysis system in the transport line.  We have tested a state-of-the-art (40 mA) proton source [7] and found a maximum $H_2^+$ current of 15 mA, with a proton-to-$H_2^+$ ratio of 1:1. Increasing the microwave power increased the proton current to 40 mA, but decreased the $H_2^+$ current.

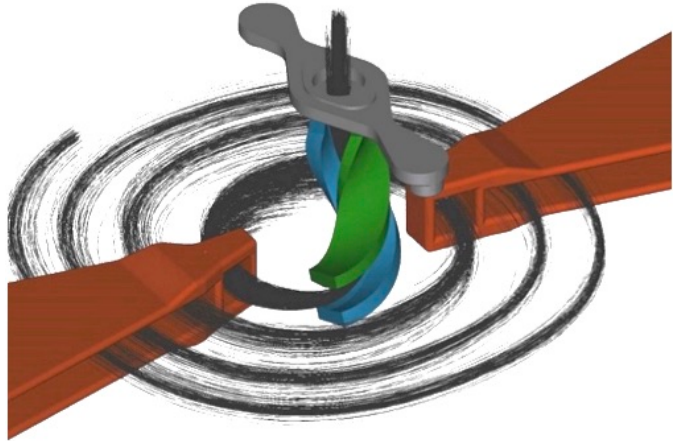

Figure 1: Schematic of the central region of a compact cyclotron. From [17]. This figure illustrates two-fold symmetry of the RF system. The IsoDAR cyclotron is designed with four-fold symmetry, there are four RF "dees."

A different source, using a filament to drive the discharge, built in the 1980's by Ehlers and Leung [8], demonstrated $H_2^+$ currents of ~80 mA. This indicates that a cooler plasma has a lower tendency to break apart the $H_2^+$ ions. A source using this technique has been assembled at MIT, called MIST-1, and is currently being commissioned [9]. It is expected to produce 30-50 mA, considerably more than needed for the cyclotron.

## B.    Bunching

Beam from the ion source emerges in a steady continuous flow, but the cyclotron will only accept a limited "phase range." Particles pass through the spiral inflector and reach the first of the accelerating electrodes (orange structures in Figure 1, called for historical reasons "dees"). These electrodes are connected to the cyclotron RF system (which operates at around 35 MHz), and so will be at a voltage that oscillates between + and - values, typically many kV. Not shown in Figure 1 are the so-called "dummy dees" which cover most of the white area and are at ground potential, presenting narrow gaps on either side of the dees. As particles pass through these gaps they are accelerated or decelerated depending on the phase of the RF at the instant each particle finds itself in a gap. In fact, only those particles passing through when the RF is within about 30° of the peak value will receive the proper acceleration to be accepted into stable orbits in the cyclotron. Hence only about 10% of a steady beam will be captured into a "bunch" and accelerated, and 90% will be lost in the first few turns. As the energy of the particles is low (extraction voltage from the ion source is typically 30 kV), these lost particles do not cause activation, but can cause thermal damage and sputtering erosion. This is commonly seen in isotope cyclotrons.

An RF buncher can be placed in the transport line to increase the beam density at the favorable phase in the RF cavity, thus improving the capture efficiency. However, the usual conventional double-gap bunchers may only increase the capture efficiency by a factor of two, i.e. from 10% to 20%.

IsoDAR uses a novel RFQ (Radio-Frequency Quadrupole) stage for bunching [10, 11, 12]. Our design, which is currently being fabricated, is expected to have bunching efficiency in excess of 80% [11]. The compact system is shown in Fig. 2. The RFQ operates at the cyclotron RF frequency and must use a split-coaxial, 4-rod configuration to resonate at such a low frequency [13]. (Typical RFQs operate at several 100 MHz.) It must be installed close to the spiral inflector to prevent loss of the high bunching factor. In addition, a small transverse focusing element (not shown) must be inserted just upstream of the spiral inflector to preserve the beam size going into the inflector. The high efficiency reduces the ion source current requirement to <7 mA, well within the expected performance of MIST-1.

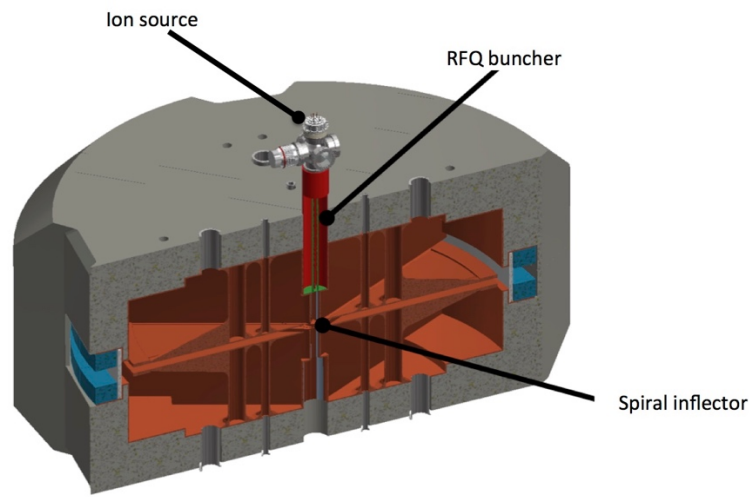

Figure 2: Components of the IsoDAR cyclotron injection system. mounted on the central axis of the cyclotron magnet. Adapted from [14].

*Running Title:* **IsoDAR Cyclotron**

## C.    Acceleration: Space-Charge and Beam Dynamics

The magnet configuration in the cyclotron consists of regions of high magnetic field (called "hills" where poles are very close to each other) and low fields ("valleys" in which there is enough space between the poles to include RF cavities), as can be seen in Figure 6. The IsoDAR magnet has a four-fold symmetry. The shape and boundaries of these regions is determined by focusing and "isochronicity" conditions, that establish that the time for a particle to make one revolution is independent of the radius of the orbit.

The electromagnetic fields in the cyclotron preserve the particle bunches by longitudinal and transverse (both horizontal and vertical) focusing forces. So, once captured, acceleration is usually very efficient, with very little beam loss. As the bunches spiral outward in the cyclotron, the separation $\Delta r$ between turns is determined by the energy gain going through the accelerating gaps, which is directly related to the voltage amplitude of the RF sine wave at these locations. Whether turns are cleanly separated depends on this $\Delta r$ and the size of the bunch, which is determined by the focusing forces from the electromagnetic fields, static and time-varying, in the cyclotron.

As the number of particles in the bunch increases, the higher Coulomb repulsion forces, referred to as "space charge," on average make the equilibrium bunch size larger, and push particles far away from the bunch center, forming so-called beam "halos." If clean turn separation is required, space charge must be taken into account very carefully.

As indicated earlier, the acceleration of $H_2^+$ instead of $H^-$ reduces space charge effects in two ways [6]. First, there are two protons for every charge, so a beam of 5 mA of $H_2^+$ contains 10 mA of protons. Secondly, the kinematic effect of space charge on bunch growth is dependent on the ion mass; the heavier mass of $H_2^+$ results in less actual growth in the beam size for a given total bunch charge. Discussed below, in Section F, is an interesting effect seen in high-current cyclotrons, where space-charge forces can actually contribute to stabilizing bunch shape, through an effect called "vortex motion." Such effects are of great importance to ensure clean turn separation, needed for efficient extraction of the beam.

## D. Extraction

When the bunch reaches the outer radius of the magnet, it is extracted and brought to the target. This can either be done using a thin electrostatic septum, defining a channel that steers the beam outside the cyclotron, or by a stripper foil.

If a septum is used, it lies between the trajectories of the final two turns, so it is important that turns are cleanly separated, otherwise beam particles will strike the septum causing activation and damage (this was a big problem with early cyclotrons).

Alternatively, for H$^-$ (or H$_2^+$) ions, one can extract the beam by placing a thin foil in the beam. Beam passing through this foil strips the two electrons from an H$^-$ ion, leaving a bare proton or, for IsoDAR, one electron is removed from the H$_2^+$ ion leaving two bare protons. As the charge to mass of the ions has changed, it is bent differently by the magnetic field. For H$^-$ the bare proton is directed outwards, away from the center of the magnet, as shown in Fig. 3.

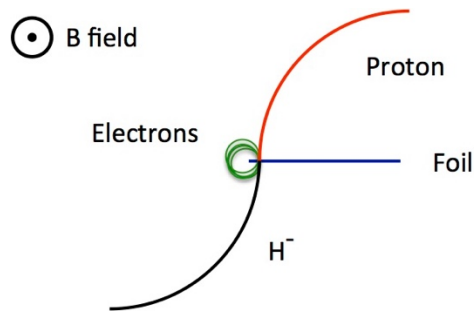

Figure 3: Schematic of the stripper foil in an H$^-$ cyclotron.

## E. Stripper Foil Extraction with H$^-$ cyclotrons

The stripper foil lifetime is the main limit to the beam current in H$^-$ cyclotrons: maximum ion source currents and bunching efficiencies, though significant, are not as important. The foils are made of carbon, around 1 micrometer (200 µg/cm$^2$) thick and mounted on an open harp, with a free edge on the beam side. Fig. 3 shows an H$^-$ ion passing through a foil at high velocity. After a few atomic layers, the ion is dissociated into a proton and two "convoy" electrons. All three particles have the same velocity, initially. If the proton has 30 MeV, the kinetic energy of each electron is 30/1836 MeV (the p/e mass ratio), or 16 keV. The foil is in a magnetic field where negative charges are bent inwards. The proton is thus bent outwards, with the same radius as the original H⁻ ion, and cleanly exits the cyclotron, but the electrons are bent inwards and their radius is smaller, also by the p/e mass ratio, so if the proton radius is 0.5 meters, the electron radius is 0.2 mm. The electrons will be bent back into the foil and will repeatedly spiral through the foil until all their energy is exhausted. The proton only makes one pass through the foil, depositing (from range/energy tables) about 2 keV, whereas the electrons deposit much more. Quantitatively, a 1 mA 30 MeV H⁻ beam deposits about 34 watts in the stripper foil, of which 94% comes from the two electrons.

Foil lifetime is determined by thermal effects and crystal dislocations. For the best carbon foils, thermal effects become important when the foil temperature exceeds 2500°C, as sublimation erodes the surface, shortening the lifetime. Above 3000°C foils are instantly vaporized. Data from a recent paper [15] indicates that 3 watts of electron power deposited in a 200 μg/cm² foil produces a temperature of 1250°C. Extrapolating to 34 watts, the T⁴ law predicts the foil will reach 2300°C, while 2 mA of 30 MeV H⁻(or 1 mA at 60 MeV) heats a foil to 2700°C. At and above these currents, foil lifetimes will be unacceptably short. It is clear that attempting to run IsoDAR-level powers cannot be done with a foil-extracted H⁻ cyclotron. The black body temperature would be about 5000°C.

## F.    Septum Extraction with $H_2^+$ Cyclotrons

Septum extraction requires clean turn separation, with the highest possible RF voltage, and a strategy for mitigating space-charge forces.

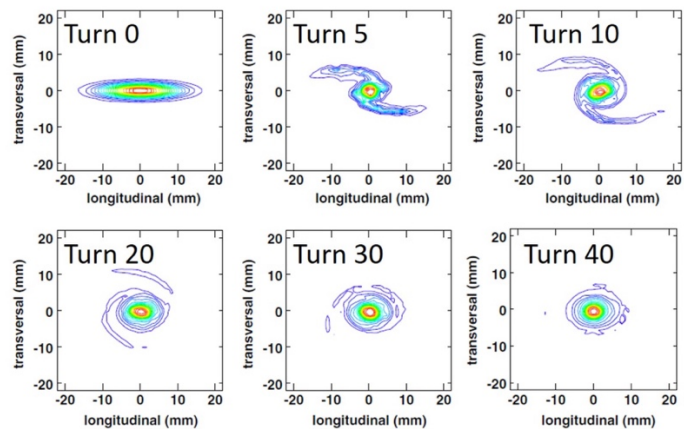

Figure 4: Evolution of a space-charge dominated bunch from injection (turn 0) to turn 40 - about midway to extraction. From [20].

Well-benchmarked simulation codes including space charge [16, 17, 18] have been used to plot the orbits and particle dynamics throughout the injection and acceleration process and verify that very little beam loss occurs between capture and the extraction radius. The objective is clean turn separation at the location of the extraction septum. With a total beam power of 600 kW, even a few parts per thousand lost on the septum can cause severe damage and activation. With these codes, halo particles that would hit the extraction septum can be traced back to early turns, and collimators can be judiciously placed to eliminate them where the particle energies are low. The strategically placed collimators and stripper foils throughout the cyclotron control activation and thermal damage to internal parts of the IsoDAR cyclotron.

Under the right conditions, vortex motion in the bunch induced by space charge forces, coupled with repulsive forces from adjacent inner and outer bunches, has been observed to actually stabilize the bunch. This effect has been observed in the high-current isochronous Injector 2 cyclotron at the Paul Scherrer Institute (PSI) [19] and has been accurately modeled (see Fig. 4) with the OPAL code [20]. The results of these calculations yield the beam distributions shown in Fig. 5 [21], with demonstrated clean turn separation at the point of the septum.

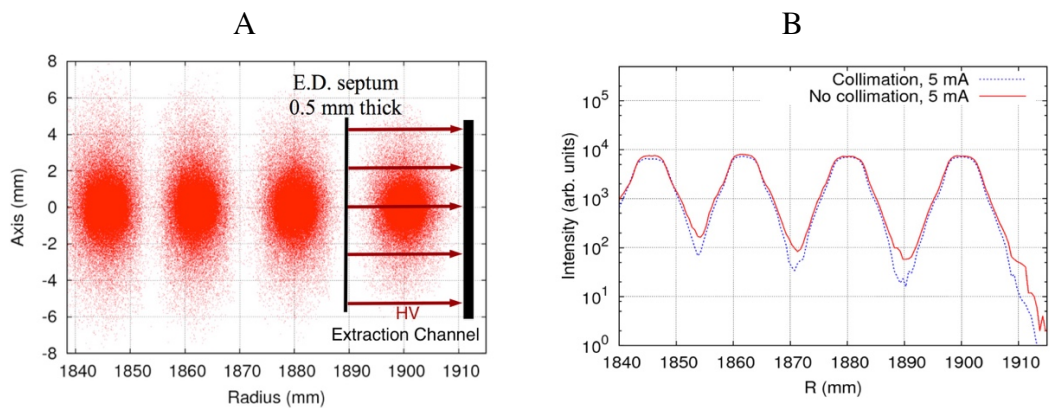

Figure 5: Simulated particle density in {y,r} plane for last few turns. Electrostatic extraction channel is shown, with field that bends beam away from cyclotron center (simulation courtesy of JJ Yang). 5(A) shows vertical beam size (mm) vs radius from center (also mm). The Electrostatic Deflection channel has a strong electric field between the plates that provides a kick to the last bunch to push it outside the cyclotron. Efficient operation requires that there be as few particles as possible in the space between the last two turns, to avoid damaging the thin septum plate. 5(B) shows the total particle count (plotted logarithmically) vs radius. The lower curve demonstrates how collimators placed close to the center of the cyclotron can help cleaning up the space between turns by absorbing halo particles. Beneficial effect of collimators is clearly seen [21].

*Running Title:* **IsoDAR Cyclotron**

## G.      Stripping Extraction with $H_2^+$ cyclotrons

Stripping foils were shown to be a current limit for $H^-$ cyclotrons because of the heat from the convoy electrons. The dynamics are different for $H_2^+$. Instead of splitting $H^-$ into one proton and two electrons, the $H_2^+$ ion is split into two protons and one electron. Hence, theoretically reducing the heat deposited by the electrons at the stripper foil by a factor of four. Because the $H_2^+$ ion is positive, the electrons are bent outwards instead of inwards, and a catcher can be placed behind the foil at an outer radius to completely suppress the electrons from re-entering the foil. Such a catcher is not possible when the electrons are bent inwards; it would interfere with the circulating beam.

If electron heating is reduced or eliminated, the limit to foil lifetime becomes crystal dislocations due to passage of the protons. We have performed an experiment, in collaboration with PSI [22], to measure the lifetime of a 79 µg/cm² foil in a 1.72 mA 72 MeV proton beam. This foil was placed in the transport line between the Injector 2 cyclotron and their main ring in an area with no magnetic fields, guaranteeing no recirculating electrons. Foil damage was seen, but only after about 60 hours of beam exposure.

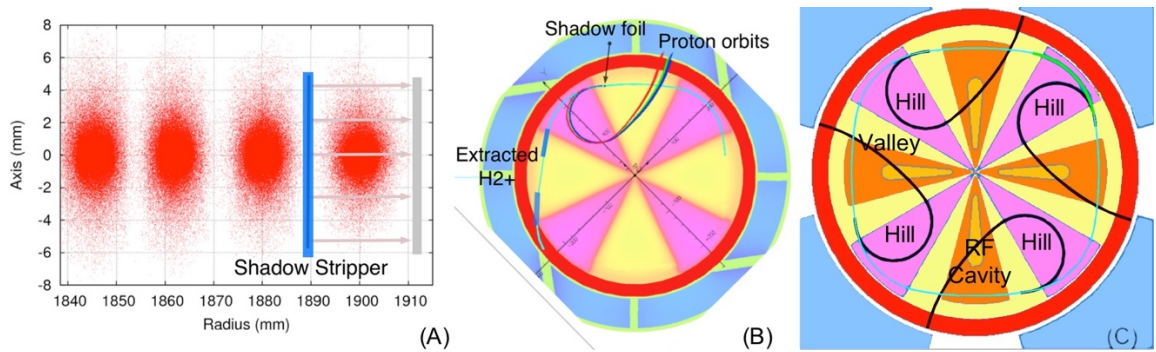

Figure 6(A): Shadow stripper protects septum from halo particles. (Underlying simulations courtesy of JJ Yang). 6(B): Orbits of protons from shadow stripper avoid the septum and exit cleanly. Different orbits correspond to changes in stripper location in hill fringe field. 6(C): 4-fold symmetry of magnetic field allows 4 locations for stripper foils to remove all beam from cyclotron via stripping. ((B) and (C) Courtesy of L. Calabretta).

In IsoDAR, the protons emerging from the stripper have a bending radius half that of the $H_2^+$ ion. Calabretta has shown (see Fig. 6 (B)), that if the stripper is placed in the correct location, the proton orbits can loop around inside the higher hill field and return into the valley region where the magnetic field is much lower, to exit cleanly from the cyclotron. In addition, if a narrow stripper foil is introduced upstream of the extraction septum, it can shadow the septum (Fig 6 (A) and (B)), and ions that would strike it are bent (as protons) to pass inside its inner edge.

Fig. 6 also shows an option where stripper foils are used for extracting all the beam from the cyclotron. As is the case with modern isotope cyclotrons, several stripper locations can be used, in our case a maximum of four, because of the four-fold symmetry of the cyclotron magnet. Using pure stripping extraction reduces the wall power needed (from 3.5 MW to 2.7 MW, see Table 1) to drive the cyclotron, because a lower RF voltage can be used since clean turn separation is not required.

## H.    Summary: IsoDAR Cyclotron Parameters

Table 1 compares the basic parameters of the IsoDAR cyclotron [23] with two leading commercial isotope cyclotrons: the IBA C30 and C70 [24]. Though the proton energy is slightly lower for the IsoDAR cyclotron (60 MeV vs 70 MeV for the IBA C70), this machine is larger and heavier because of the higher magnetic rigidity of the $H_2^+$ ion accelerated, but these penalties are outweighed by the benefits.  IsoDAR delivers roughly ten times the current of the C70 but requires significantly less than ten times the wall plug power, and for less than three times the footprint.

Table 1:  Comparison of IsoDAR with IBA commercial isotope cyclotrons

| Parameter | IsoDAR | IBA C30 | IBA C70 |
|---|---|---|---|
| Ion species accelerated | $H_2^+$ | $H^-$ | $H^-$ |
| Maximum energy (MeV/nucleon) | 60 | 30 | 70 |
| Proton beam current (mA) | 10 | 1.2 | 0.75 |
| Available beam power (kW) | 600 | 36 | 52 |
| Pole radius (m) | 1.99 | 0.91 | 1.24 |
| Outer diameter (m) | 6.2 | 3 | 4 |
| Iron weight (metric tons) | 450 | 50 | 140 |
| Electric power required (MW) | 3.5 or 2.7 | 0.15 | 0.5 |

# III   ISOTOPE APPLICATIONS

With the factor of 10 increase in beam current, the benefits of an IsoDAR-class cyclotron are higher production rates for lower cross section isotopes and efficient production of larger amounts of long-lived isotopes. This will require development of targets that can take advantage of the high powers. We address the flexibility of a Q/A = 0.5 cyclotron to accelerate a wide variety of ions, and, finally, examples are given of the yields possible for two isotopes in high demand at present.

## A. Targetry and Beam Power Management

Due to the very high beam power of the IsoDAR cyclotron, state of the art facilities would be required for handling and processing of such a high level of isotope production, as well as for ensuring radiation safety. While this an important challenge, it is not a particularly novel one and should be overcome with feasible extensions of existing technologies.

Targets are currently designed for a maximum of a few kW. Recent developments [25] extend this into a few 10's of kW. The full 600 kW is considerably beyond the present state of the art.

As power density is a key parameter, development activities probably should concentrate on larger targets and capabilities for spreading beam over larger areas.  In addition, focus should be on more efficient cooling and heat-transport designs.  These must be coupled with metallurgy and chemistry of suitable isotope source materials.  Operating at higher temperatures should also be a parameter investigated, for instance using liquid metal targets where this might be possible, and high-temperature coolants.

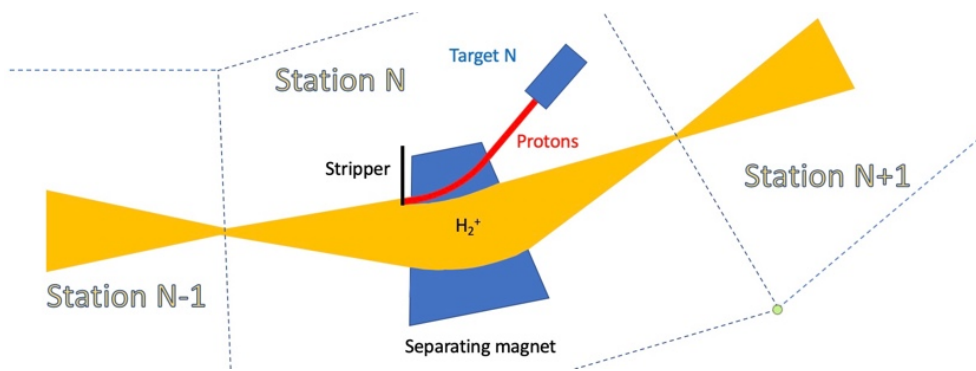

Figure 7: Technique for sharing beam between many targets using sequential stations.

While having such a powerful beam presents an opportunity for development of more heat-tolerant targets, strategies exist for splitting the beam amongst several targets. For IsoDAR, up to four stations can be used with internal strippers, though experience with H$^-$ cyclotrons indicates that tuning for more than two at a time may prove difficult.

A technique has been proposed [26] using an extracted H$_2^+$ beam, that consists of a modular transport line with stations where a small amount of beam is peeled off and directed to a target. As shown in Fig. 7 each station would have a separating magnet. Just upstream of the magnet, a stripper foil is inserted into the edge of the broadened H$_2^+$ beam, producing a proton fraction that is directed to a target. The remaining H$_2^+$ is refocused and sent to the next station. In this way adjustable amounts of protons can be sent to many targets.

Key for developing higher power targets is to spread the beam over a larger area, thus keeping the power density on the target to within limits for cooling. In doing so it is important to ensure that beam uniformity over the area of the target is maintained, and that there are no hot spots on the target. This can be done with beam line magnets, including quadrupole and higher-order (sextupole and octupole) focusing elements, and also with raster or wobbler magnets that rapidly move the beam spot over the face of the target. Both techniques have been used successfully for ensuring uniform irradiation of targets.

The beam power available from the cyclotron has the potential of causing damage to internal and external components. In the event of unexpected beam deviations, appropriately placed radiation monitors will detect increase in beam loss and enable shutting the beam off before damage can occur.

Our designs and simulations indicate that the full beam power can be extracted without untoward beam losses, however reaching this level of performance will require very careful tuning and commissioning. To do this, it will be necessary to run at lower powers, and slowly ramp up the power as losses are reduced.

As indicated earlier, space charge is a very important element in the beam dynamics. To reduce total power, then, it is best to reduce the duty factor rather than reducing the current level from the source. A chopper has been built into the short transport line between the source and the RFQ, so time structure can be introduced into beam injected into the RFQ. This chopper can turn the beam on or off in under a microsecond and is adjustable to provide a beam duty factor between 0 and 100%. In this way the beam power can be adjusted without changing the number of particles in an RF bunch, preserving the space charge effects in the bunch.

This adjustment in duty factor can also be used productively for developing higher power targets, to continuously increase the power on a target up to its design limit. One should note that the chopper timing needs to be coordinated with any beam sweeping over the target, and with thermal time constants in the target heat-handling designs.

## B.      Beam Species Flexibility

A cyclotron designed to accelerate $H_2^+$ can, with only minor tuning changes, accept any ion with a charge-to-mass ratio of 0.5. So, deuterons, $He^{++}$, $C^{6+}$ or other like ions could be accelerated to the same energy-per-nucleon (60 MeV) as $H_2^+$. Slight tuning changes are needed because of the small proton/neutron and nuclear binding mass differences; very small adjustments to the magnetic field are needed to preserve isochronicity for each ion species. Such adjustments could be done with trim coils placed in the valley regions of the cyclotron magnet. For ions other than $H_2^+$, the ion source would need to provide the fully-stripped ions injected. While it is difficult to remove all the electrons from helium and carbon atoms, a commercially-available Electron Cyclotron Resonance (ECR) source, the PK-ISIS unit from Pantechnik [27], delivers 2.4 mA of alpha particles and 50 µA of $C^{6+}$. Beyond the ion source, the transport, bunching, injection and acceleration of these ions does not differ from $H_2^+$. As these ions are fully stripped, foils will not change the charge-state of the ion, so conventional septum extraction must be used. Expected beam-on-target for these ions would be about 1 mA for alpha particles (240 MeV, at 120 kW of beam power) or 30 µA of carbon (3.6 kW, 720 MeV of total energy).

Deuteron beams of 5 mA would be indistinguishable, as far as the accelerator is concerned, from the planned $H_2^+$ beams (except, again, that foils cannot be used for extraction). This current level would be easily obtained from a standard proton source using deuterium instead of hydrogen as the source gas. Because of the prevalence of breakup of the deuteron in the target and production of beam-energy neutrons, the limit on deuteron beam current would probably come from the facility shielding emplaced. In principle, beam power on target could reach 600 kW.

These beam species, and power levels, provide for totally new areas of research in isotope production. Regions inaccessible at present would open up, and while most isotopes directly produced would have short half-lives, decay chains could yield new or existing isotopes that could prove interesting and economical for medical or other applications.

## C.     Example: A $^{68}$Ge/ $^{68}$Ga Generator for diagnostic imaging

The high power of the IsoDAR cyclotron opens the possibility of highly efficient production of the $^{68}$Ge/ $^{68}$Ga generator. This generator has many advantages in a clinical setting and improving its accessibility and reducing production costs can have a very large impact on nuclear medicine. $^{68}$Ge decays with a half-life of 270 days to $^{68}$Ga, which is a positron emitter, finding increasing application in PET imaging.

Generators (often referred to as "cows") offer great advantages for nuclear medicine studies, in that the imaging isotope is available without an on-site accelerator. A long-lived parent is produced in an accelerator, or reactor, and is shipped to the use site. The short-lived daughter is "milked" from the source as required, this short-lived isotope is used in the diagnostic study.

The usefulness of a generator is related to the half-life of the parent. If one compares the $^{68}$Ge/ $^{68}$Ga generator with the widely-used $^{99}$Mo/ $^{99m}$Tc generator, the $^{99}$Mo lifetime is 66 hours, so such generators have a useful lifetime of a week or two, whereas the 270-day half-life of $^{68}$Ge provides a much longer shelf life, typically one year.

Because of the long parent half-life, production of a viable $^{68}$Ge/$^{68}$Ga generator requires many hours of cyclotron time, leading to high costs and scarce availability. The IsoDAR cyclotron, with its factor of 10 higher beam current, immediately increases the production rate by a factor of 10. In addition, however, we note that the higher energy of the proton beam (60 MeV) can almost double again the production yield.

Natural gallium, the target material, has two isotopes, $^{69}$Ga (60% abundance) and $^{71}$Ga (40%). Both these isotopes can produce $^{68}$Ge: $^{71}$Ga(p,4n)$^{68}$Ge, and $^{69}$Ga(p,2n)$^{68}$Ge. Both are compound nucleus reactions, with approximately equal cross sections (around 150 mb), the first peaks at a proton energy of around 50 MeV, the second at about 25 MeV. Both have excitation function widths of about ± 5 MeV. So, bringing a 60 MeV proton beam into a thick target of gallium will first make use of the heavier isotope, and as the protons lose energy will produce the $^{68}$Ge from the lighter isotope. Other Ge isotopes produced in the target have substantially lower half-lives than $^{68}$Ge; the longer-lived ones are: $^{71}$Ge (11 days), $^{69}$Ge (39 hours) and $^{66}$Ge (2.2 hours). $^{71}$Ge and $^{98}$Ge decay to stable Ga isotopes so do not contaminate the generator and waiting a day before processing the target adequately removes any $^{66}$Ga from the generator.

The 10-mA intensity of the proton beam from the IsoDAR cyclotron could produce, assuming the above cross sections, approximately 50 Curies of the $^{68}$Ge parent in a week of running. This could yield a very large number of generators, which, with a year or more useful lifetime, could greatly reduce the dependence on a rapid supply chain for distribution of the generator.

This production rate assumes that all of the 600 kW of available beam power can be deposited on production targets. If stripping extraction is used, and all four ports used simultaneously, each target would need to absorb 150 kW. As gallium has a low melting point, the metal in the target would undoubtedly be in liquid form. Concepts for high-power liquid gallium targets have been developed, the current limit is around 50 kW [25]. Extending this to 150 kW or higher will require further development efforts.

### D.    Example: Production of $^{225}$Ac: an α emitter for targeted radiotherapy

Alpha-emitting isotopes are in high demand for therapeutic applications. The short range of alpha particles, and the high LET (Linear Energy Transfer) of the stopping alpha lead to extremely effective cell killing. One of the most effective isotopes is Actinium 225, with a 9.9-day half-life. It is the parent of a chain that includes four alpha emitters ending up with stable $^{209}$Bi. The four alpha particles at the site of the original $^{225}$Ac all contribute to the radiation damage to the cells within a radius of about 50 μ-meters of the decaying nucleus. Reference [28] outlines the development of this radioisotope for clinical applications.

The initial source of $^{225}$Ac arose from the chemical separation of $^{229}$Th from $^{233}$U. For this process, the sophisticated hot-chemistry resources at Oak Ridge and Karlsruhe were used. Alpha decay of $^{229}$Th (8000 year half-life) could yield small quantities of $^{225}$Ra that then beta decayed (with a 14-day half-life) to $^{225}$Ac. Though very complex, this process did yield small quantities of $^{225}$Ac, sufficient for some highly successful clinical studies. Another production method is proton irradiation of $^{226}$Ra, that yields $^{225}$Ac via a (p,2n) reaction. However, isolating sufficient $^{226}$Ra for the cyclotron targets involves a process almost as complex as the one described above.

A more promising possibility arose from studies at Los Alamos, where thick targets of natural thorium were bombarded with 200 MeV protons. In these experiments, researchers demonstrated that $^{225}$Ac can be produced with acceptable efficiency [29]. Cumulative cross sections were measured, from 200 MeV (15 mb) to 50 MeV (5 mb). Their publication states that use of BLIP (Brookhaven) and LANSCE (LANL) at 100 μA for production of $^{225}$Ac could increase the world supply by a factor of 60. Increasing the current from 100 μA to 10 mA increases this number by another factor of 100.

We estimate the IsoDAR production rate from a thorium target to be around 7.4 gigabecquerel (200 mCi) per hour. Thus, in 5 hours, we match the current yearly production. This application will require two technical advances. First, the development of high-power thorium targets. However, thorium has a high melting point, so it can withstand considerable heating. A rotating target configuration might provide a good path to the high powers needed. Second, appropriate separation processes to extract the $^{225}$Ac from the bombarded target must be devised. This will be complex due to radioactivity in the target. These are solvable problems that are motivated by the game-changing quantities of $^{225}$Ac that IsoDAR can provide.

## IV.   Conclusion

Meeting the beam-current requirement for the IsoDAR Cyclotron, to satisfy its mission as a driver for a neutrino source, has led to the development of the capability to produce high current proton beams. This places it above all existing cyclotrons. Achieving these currents has required innovative developments in ion sources, bunching and injection, capture, acceleration and extraction of the ions in a highly optimized cyclotron design. The IsoDAR team has made good progress towards demonstrating the expected performance.

The intensity increase improves commercial and clinical viability of difficult-to-produce radioisotopes, such as $^{225}$Ac and the long-lived $^{68}$Ge/ $^{68}$Ga PET generator. As newly emerging isotopes are investigated, this flexible, enhanced performance technology will allow for more efficient evaluation of new candidates. Isotope target development to fully utilize the available beam power will be a significant challenge. However, having the very high beam power available provides an effective test bed for target development. The capability of such high production rates of rare isotopes is a strong incentive to fully utilize the power of a high-power cyclotron. In addition, techniques for beam splitting which are unique to $H_2^+$ could make efficient use of lower-power targets. The powerful capabilities of the IsoDAR cyclotron including novel target development, multiple beam species flexibility, and massive production rates of rare medical isotopes demonstrate that it is the next step in high-power cyclotron technology.

# V.    DECLARATIONS

## A.    Consent for public

Not applicable

## B.    Ethics Approval and Consent to Participate

Not applicable

## C.    Availability of data and Material

Data sharing not applicable to this article as no datasets were generated or analyzed during the current study.

## D.    Funding

This work supported by the US National Science Foundation, Grant NSF-PHY-1505858.

## E.    Authors Contributions

JA, JC and LW developed contacts with the isotope industry. JC leads the cyclotron development effort. RB and JA wrote the first drafts of the manuscript, all contributed to its editing.

## F.    Acknowledgements

We are grateful for fruitful discussions with colleagues in the nuclear medicine and radioisotope fields; Richard Johnson, University of British Columbia; Jacob Hooker, Mass General; Jerry Peterson, University of Colorado, Boulder; and Suzanne Lapi, University of Alabama, Birmingham. We especially thank our collaborators Luciano Calabretta, who originated the idea of use of $H_2^+$; Andreas Adelmann, a leader in adapting the OPAL simulation for use in cyclotrons; and the IsoDAR cyclotron development team, consisting of scientists and engineers from MIT, the Paul Scherrer Institut, Laboratori Nazionali del Sud, Italy, and IBA Radiopharma

*Running Title:*  **IsoDAR Cyclotron**

Solutions. This work was supported by the US National Science Foundation, Grant NSF-PHY-1505858.

## G. Competing Interests

The authors declare that they have no competing interests.

*Running Title:* IsoDAR Cyclotron

*Running Title:*  IsoDAR Cyclotron